\documentclass{ws-procs975x65}

\begin{document}

\title{DEWITT-SCHWINGER RENORMALIZATION OF $\langle\phi^2\rangle$ IN $d$ DIMENSIONS}

\author{ROBERT T. THOMPSON$^{\dag}$ and JOS\'E P. S. LEMOS$^{\ddag}$}

\address{Centro Multidisciplinar de Astrof\'{\i}sica - CENTRA \\
Departamento de F\'{\i}sica, Instituto Superior T\'ecnico - IST,\\
Universidade T\'ecnica de Lisboa - UTL,
Av. Rovisco Pais 1, 1049-001 Lisboa, Portugal\\
E-mail: $^{\dag}$robert@cosmos.phy.tufts.edu, $^{\ddag}$joselemos@ist.utl.pt}

\begin{abstract} 
A compact expression
for the DeWitt-Schwinger renormalization terms suitable for use in
even-dimensional space-times is derived.  This formula should be useful
for calculations of $\langle\phi^2(x)\rangle$ and $\langle
T_{\mu\nu}(x)\rangle$ in even dimensions.
\end{abstract}

\bodymatter

\section{Introduction}
A major impediment to using semi-classical general relativity is calculating the renormalized
expectation value of the stress tensor.  Properly renormalized values
for $\langle\phi^2\rangle$ and $\langle T_{\mu\nu}\rangle$ provide
information on particle production and spontaneous symmetry breaking,
and are also required to calculate backreaction.  Since in general
relativity energy density is itself a source of curvature, great care must be taken in deciding what 
may be dismissed as `unphysical'.  Fortunately, there are several generally
accepted renormalization schemes for curved space-times
\cite{Birrell:1982ix}.  Our purpose is to present a compact
formula for the renormalization terms that may be applied to
$\langle\phi^2\rangle$ and $\langle T_{\mu\nu}\rangle$ calculations in arbitrary black hole space-times of even dimension.

\section{Connection to Green's Functions}
Calculating $\langle T_{\mu\nu}\rangle$ for a general d-dimensional black hole space-time is difficult.  For a scalar field, $ T_{\mu\nu} \propto \phi^2$ and its derivatives, so we start with the simpler problem of calculating $\langle\phi^2\rangle = \langle H|\phi^2|H\rangle$, where $|H\rangle$ is the Hartle-Hawking vacuum.  Note that $\langle\phi^2\rangle$ is the coincidence limit of the two point function $\langle\phi^2\rangle = \lim_{x\to x'}\langle\phi(x)\phi(x')\rangle$, and so may be expressed in terms of Green's functions.  In particular, the Feynman Green's function is related to the time ordered propagator, $iG_{\rm F}(x,x')=\langle T\left(\phi(x)\phi(x')\right)\rangle$.  A Wick rotation allows us to work in Euclidean space where $G_{\rm F}(t,x;t',x')=-iG_{\rm E}(i\tau,x;i\tau',x')$.  The Euclidean Green's function, $G_{\rm E}$, now obeys
\begin{equation} \label{Eq:Green}
\left(\square_{\rm{E}}- m^2 - \xi R(x)\right)G_{\rm E}(x,x')=
-|g(x)|^{-1/2}\delta^d(x-x'),
\end{equation}
where $\square_{\rm{E}}$ is the Laplace-Beltrami operator in $d$-dimensional curved Euclidean space. To solve for $G_{\rm E}$, start with the Euclidean
metric for a static space-time in $d$ dimensions with line element 
\begin{equation} \label{Eq:Metric}
 ds^2 = f(r)d\tau^2 + f^{-1}(r)dr^2 + r^2d\Omega^2.
\end{equation}
Here $\tau$ is the Euclidean time, $\tau = -it$, $r$ is a radial coordinate, and $\Omega$ represents a $(d-2)$-dimensional angular space.  The only restriction for this method is that the metric must be diagonal. If the scalar field is at temperature $T$, then the Green's function is periodic in $\tau-\tau'$ with period $T^{-1}$.  Assuming a separation of variables, standard Green's function techniques lead to the formal solution
\begin{equation} 
\label{Eq:GreenFuncNonZeroTemp}
 G_{\rm E}({x},{x'}) = \frac{\kappa}{2\pi}
\sum_{n=-\infty}^{\infty} e^{i\kappa\varepsilon n}\sum_{\ell}
\sum_{\{\mu_j\}}Y_{\ell,\{\mu_j\}}(\Omega)Y^*_{\ell,\{\mu_j\}}(\Omega') 
\chi_{n\ell}(r,r'),
\end{equation}
where $\kappa = 2\pi T$, and $Y_{\ell,\{\mu_j\}}(\Omega)$ are eigenfunctions of the Helmholtz equation obtained from the from the angular part of Eq.\ (\ref{Eq:Green}).  For black holes with spherical topology these are equivalent to the set of hyperspherical harmonics.  The radial function $\chi_{n\ell}(r,r')$ obeys a complicated differential equation obtained by putting the above expression into Eq.\ (\ref{Eq:Green}).  This expression is divergent in the sum over $n$.

\section{DeWitt-Schwinger Renormalization}
The $\langle \phi^2\rangle$ computation has been reduced to computating the coincidence limit of the Green's function -- a divergent quantity.  To assign physical meaning to $\langle\phi^2\rangle$ it must be
rendered finite via some renormalization process, and the standard approach is to renormalize
the expression for $G_{\rm E}({x},{x'})$ via Christensen's point splitting
method applied to the DeWitt-Schwinger expansion of the
propagator
\cite{Christensen:1976vb,Christensen:1978yd,DeWitt:1975ys}.
In $d$ dimensions, the adiabatic DeWitt-Schwinger expansion of the
Euclidean propagator is \cite{Christensen:1978yd}
\begin{equation} \label{Eq:SchwingDeWitt}
G_{\rm E}^{\rm DS}({x},{x'}) = \frac{\pi \triangle^{1/2}}{(4\pi
i)^{d/2}} \sum_{k= 0}^{\infty} a_k({x},{x'})
\left(-\frac{\partial}{\partial m^2}\right)^k 
\left(-\frac{z}{2im^2}
\right)^{1-d/2} H^{(2)}_{d/2-1}(z).
\end{equation}
Equation (\ref{Eq:SchwingDeWitt}) introduces
several new variables.
Let $s({x},{x'})$ be the geodesic distance between
${x}$ and ${x'}$, then define
$2\sigma({x},{x'}) = s^2({x},{x'})$ and $z^2=-2m^2\sigma({x},{x'})$.  
The
$a_k({x},{x'})$ are called
DeWitt coefficients, and $H^{(2)}_{\nu}(z)$ is a
Hankel function of the second kind. Lastly,
$\triangle({x},{x'}) =
\sqrt{g({x})}D({x},{x'})\sqrt{g({x'})}$ is
the Van Vleck--Morette determinant, where
$g({x})=\det(g_{\mu\nu}({x}))$ and
$D({x},{x'})=\det(-\sigma_{;\mu\nu'})$.
Using the derivative properties of Bessel functions, noting that $z=i|z|$ is purely imaginary in Euclidean space, and defining $\nu=d/2-1-k$, Eq.\ (\ref{Eq:SchwingDeWitt}) can be written as
\begin{equation} \label{Eq:SDWIntermediate}
G_{\rm E}^{\rm DS}({x},{x'}) = \frac{-2i
\triangle^{1/2}}{(4\pi)^{d/2}} \sum_{k= 0} a_k({x},{x'})(2m^2)^{\nu}
|z|^{-\nu}\Big[(-1)^{\nu}\pi I_{\nu}(|z|)+i K_{\nu}(|z|)\Big].
\end{equation}

The DeWitt-Schwinger expansion is a WKB expansion of the Euclidean propagator for a generic space-time when the point separation is small.  For a particular space-time, this procedure does not give the correct results for the
Green's function with finite point separation because it ignores global space-time properties that determine the Green's function -- such as the effective potential around a black hole -- but it should reproduce the same divergent terms in the coincidence limit. Therefore, if the divergent terms of the DeWitt-Schwinger expansion
can be isolated, then subtracting these terms from $G_{\rm E}({x},{x'})$ will make it finite as $x\rightarrow x'$. Since we are working in Euclideanized space the physical
renormalization terms come from the real part of Eq.\ (\ref{Eq:SDWIntermediate}).  The asymptotic
behavior of $K_{\nu}(|z|)$ as $z\to 0$ implies that only terms with $\nu\geq 0$ contribute divergences in the coincidence limit, so
\begin{equation} \label{Eq:GdivTerms} G_{\rm div}({x},{x'})
= \frac{2 \triangle^{1/2}}{(4\pi)^{d/2}} \sum_{k= 0}^{k_d}
a_k({x},{x'}) (2m^2)^{\nu} |z|^{-\nu} K_{\nu}(|z|).
\end{equation}

To renormalize $G_{\rm E}$, Eq.\ (\ref{Eq:GdivTerms}) must be made commensurate with Eq.\ (\ref{Eq:GreenFuncNonZeroTemp}).  We have shown \cite{Thompson:2008bk} that an integral representation of $K_{\nu}(z)$ for small $z$ and integer-valued $\nu$ is
\begin{equation} \label{Eq:IntegralRepresentation}
K_{\nu}(z) =
\frac{(-1)^{\nu}\sqrt{\pi}}{\Gamma(\nu+\tfrac{1}{2})}
\left(\frac{z}{2}\right)^{\nu}
\int_0^{\infty}dt\cos(zt)(t^2+1)^{\nu-1/2}.
\end{equation}
Changing of variables and using the Plana sum formula to convert the integral to a sum, the renormalization terms for the $d$-dimensional space-time of Eq.\ (\ref{Eq:Metric}) are \cite{Thompson:2008bk}
\begin{multline} \label{Eq:GdivNonZero}
  G_{\mathrm{div}}(x,x') = \frac{2}{(4\pi)^{d/2}}\sum_{k=0}^{k_d}
\Bigg\{\frac{\left[a_k\right]\kappa\sqrt{\pi}}{(-f)^{\nu}
\Gamma(\nu+\frac{1}{2})} 
\Bigg[\sum_{n=1}^{\infty}\cos(\kappa\varepsilon n)
\left(\kappa^2n^2+m^2f\right)^{\nu-\frac{1}{2}} - 
\frac{1}{2}(\kappa^2+m^2f)^{\nu-\frac{1}{2}} \\ - i\int_0^{\infty}
\frac{dt}{e^{2\pi
t}-1}\left\{\left[(1+it)^2\kappa^2+m^2f\right]^{\nu-1/2} -
\left[(1-it)^2\kappa^2+m^2f\right]^{\nu-1/2}
\right\} \\ + (m^2f)^{\nu-\frac{1}{2}}\phantom{}_2F_1
\left(\frac{1}{2},\frac{1}{2}-\nu,\frac{3}{2},-\frac{\kappa^2}{m^2f}\right)
\Bigg]+ [a_k]E_{\nu} + \sum_{n=1}^{\nu}\sum_{p=1}^{2n}\sum_{j=0}^{p}
\frac{2^{2n-1}(-m^2)^{\nu-n}\Gamma(n)}{\Gamma(\nu-n+1)}
\frac{a^j_k\Delta^{1/2}_{p-j}}{(\sigma^{\rho}\sigma_{\rho})^n}  \Bigg\}
\end{multline}
for a scalar field at nonzero temperature $T>0$.
In this expression the $E_{\nu}$ are terms depending on the metric function $f$ and have been tabulated elsewhere \cite{Thompson:2008bk}, while $a_k^m$ and $\Delta^{1/2}_{m}$ represent the $m^{th}$ term in an expansion in powers of $\sigma^{\rho}$.  This expression generalizes previously known four-dimensional results \cite{Anderson:1990jh}. The corresponding renormalization terms for a scalar field at zero temperature $T=0$ are similarly found \cite{Thompson:2008bk}.

\section{Discussion}
Semi-classical general relativity requires calculation of $\langle T_{\mu\nu}\rangle_{\rm{ren}}$ in complicated -- possibly higher dimensional -- space-times.  The first step in calculating $\langle T_{\mu\nu}\rangle_{\rm{ren}}$ for a scalar field is calculating $\langle \phi^2\rangle_{\rm{ren}}$.  We have presented a compact expression for the renormalization terms for $\langle \phi^2\rangle$ in even dimensional, static, black hole space-times.
\section*{Acknowledgments}
This work partially funded by FCT, through Project No. CERN/FP/109276/2009 

\bibliographystyle{ws-procs975x65}

\end{document}